\documentclass[aps,onecolumn,nofootinbib,superscriptaddress,floatfix]{revtex4}

\usepackage{color,amsmath,amssymb,graphicx,latexsym,subfigure}

\usepackage{multirow}
\usepackage{ulem}
\usepackage{color}
\usepackage{float}
\usepackage{booktabs}
\usepackage{threeparttable}
\usepackage[colorlinks,linkcolor=green, anchorcolor=green,citecolor=green]{hyperref}
\usepackage{ulem,soul}
\usepackage{xcolor}

\newcommand{\be}{\begin{equation}}
\newcommand{\ee}{\end{equation}}
\newcommand{\ba}{\begin{eqnarray}}
\newcommand{\ea}{\end{eqnarray}}

\newcommand{\gsim}{\mathrel{\hbox{\rlap{\lower.55ex \hbox {$\sim$}}
			\kern-.3em \raise.4ex \hbox{$>$}}}}
\newcommand{\lsim}{\mathrel{\hbox{\rlap{\lower.55ex \hbox {$\sim$}}
			\kern-.3em \raise.4ex \hbox{$<$}}}}

\makeatletter
\def\section{\@startsection {section}{1}{\z@}%
    {-3.5ex \@plus -1ex \@minus -.2ex}%
    {2.3ex \@plus.2ex}%
    {\normalfont\bfseries\boldmath\rightskip\z@}}
\makeatother

\hypersetup{colorlinks=true,
	breaklinks=true,
	pdfstartview=Fit,
	linkcolor=blue,
	citecolor=blue,
	urlcolor=blue}

\begin{document}

\title{A Focused Review of Quintom Cosmology: From Quintom Dark Energy to Quintom Bounce}

\author{Tao-tao Qiu} \email{qiutt@hust.edu.cn}
\affiliation{School of Physics, Huazhong University of Science and Technology, Wuhan, 430074, China}

\author{Yifu Cai} \email{yifucai@ustc.edu.cn}
\affiliation{Department of Astronomy, School of Physical Sciences,
 University of Science and Technology of China, Hefei 230026, China}
\affiliation{CAS Key Laboratory for Research in Galaxies and Cosmology,
 School of Astronomy and Space Science, University of Science and Technology of China, Hefei 230026, China}

\author{Yang Liu} \email{liuy92@ihep.ac.cn}
\affiliation{Key Laboratory of Particle Astrophysics, Institute of High Energy Physics, Chinese Academy of Sciences, Beijing 100049, China}

\author{Si-Yu Li} \email{lisy@ihep.ac.cn}
\affiliation{Key Laboratory of Particle Astrophysics, Institute of High Energy Physics, Chinese Academy of Sciences, Beijing 100049, China}



\author{Jarah Evslin} \email{jarah@impcas.ac.cn}
\affiliation{Institute of Modern Physics, NanChangLu 509, Lanzhou 730000, China}
\affiliation{University of the Chinese Academy of Sciences, YuQuanLu 19A, Beijing 100049, China}
 
\author{Xinmin Zhang} \email{xmzhang@ihep.ac.cn}
\affiliation{Theoretical Physics Division, Institute of High Energy Physics, Chinese Academy of Sciences, 19B Yuquan Road, Shijingshan District, Beijing 100049, China}
\affiliation{School of Nuclear Science and Technology, University of Chinese Academy of Sciences, Beijing,101408, China}

\begin{abstract}

The recently released data of DESI DR2 favors a dynamical dark energy theory, with the equation of state crossing the cosmological constant boundary $w=-1$. In this paper, we briefly review quintom cosmology, especially the quintom bounce. We will give three examples of a quintom bounce and one example of a cyclic universe with quintom matter. 

\end{abstract}

\maketitle


The accelerated expansion of the Universe was first discovered in 1998 through observations of high-redshift Type Ia supernovae (SNe)~\cite{Riess_1998, Perlmutter_1999}, and has since been robustly confirmed by cosmic microwave background (CMB) anisotropies, baryon acoustic oscillations (BAO), and large-scale structure data. Within the standard cosmological model, this acceleration is attributed to a cosmological constant $\Lambda$, yielding the highly successful $\Lambda$CDM paradigm for past 20 years.

Recent results from Dark Energy Spectroscopic Instrument (DESI) Data Release 2, when combined with CMB and supernova data, provide compelling evidence that dark energy may be dynamical in nature~\cite{DESI:2025zgx, DESI:2025wyn}. The joint analysis reveals a statistically significant time evolution of the dark energy equation of state (EoS), $w$, with the DESI DR2+CMB+DESY5 dataset excluding the $\Lambda$CDM model at the $4.2\sigma$ level. Notably, the reconstructed $w$ crosses the cosmological constant boundary at $w = -1$, a distinctive signature of the \emph{Quintom} scenario~\cite{Feng:2004ad}.

To quantify the statistical support for the Quintom scenario, we perform an independent cosmological analysis using the same combined DESI DR2+CMB+DESY5 dataset. Here, CMB includes the full Planck temperature (TT), polarization (EE), and cross (TE) power spectra (using \texttt{simall}, \texttt{Commander} for $\ell < 30$, and \texttt{CamSpec} for $\ell \geq 30$), supplemented by the Planck+ACT DR6 CMB lensing likelihood. Using the \texttt{Cobaya} inference framework with Metropolis–Hastings MCMC sampling, interfaced with \texttt{CAMB} for theory predictions, we adopt the CPL parametrization $(w_0, w_a)$ \cite{Chevallier:2000qy, Linder:2002et} for dark energy, and the standard six-parameter base ($\omega_b$, $\omega_c$, $100\theta_{\rm MC}$, $\ln(10^{10}A_s)$, $n_s$, $\tau$). Chains are deemed converged when the Gelman–Rubin statistic satisfies $R - 1 < 0.01$, and posterior summaries are generated using the \texttt{GetDist} package.

\begin{figure}[htbp]
    \centering
        \includegraphics[width=0.6\textwidth]{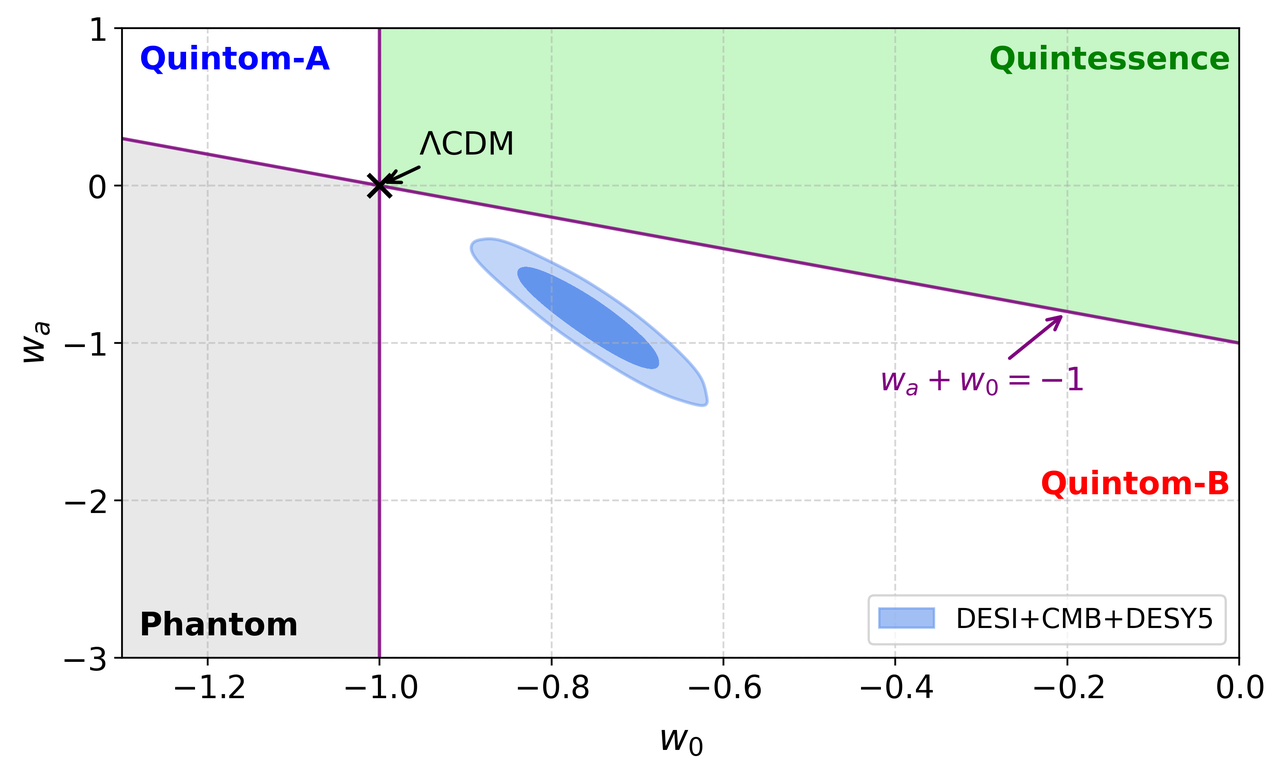}
    \caption{Results for the posterior distributions of $w_0$ and $w_a$, from fits of the $w_0w_aCDM$ model to DESI in combination with CMB and DESY5 datasets as labelled. Quintom-A: $w$ evolves from Quintessence to Phantom; Quintom-B: from Phantom to Quintessence.}
    \label{desi_w0wa}
\end{figure}

Our result yields $\Delta \chi_{\text{MAP}}^2 = -21.2$ relative to the $\Lambda$CDM model, corresponding to a $4.22\,\sigma$ deviation. Moreover, we find a posterior probability of $99.997439\%$ for the Quintom-B region defined by $w_0 > -1$ and $w_0 + w_a < -1$, equivalent to a significance of approximately $4.05\,\sigma$. As shown in Fig.~\ref{desi_w0wa}, our results are fully consistent with the DESI findings~\cite{DESI:2025zgx} and very strongly support the Quintom scenario.\footnote{Since in Quintom-B scenario, $w<-1$ stays only for a short period of time, the null energy condition can still be satisfied for the whole universe \cite{Qiu:2007fd, Caldwell:2025inn}.}

The Quintom scenario was first proposed in April of 2004~\cite{Feng:2004ad}. The Quintom theory of dark energy differs from Quintessence and the Phantom, and predicts the EoS of dark energy crossing\footnote{This is also called phantom divide crossing in the literature.} the cosmological constant boundary at $w = -1$. In general, the basic single-field models or single perfect fluid models can not exhibit Quintom behavior due to the No-Go theorem \cite{Xia:2007km}. For a detailed description of the Quintom model, please refer to Refs.~\cite{Guo:2004fq, Zhao:2005vj, Cai:2009zp, Qiu:2010ux, Cai:2025mas}.

In 2007, we considered the applications of Quintom scenarios on the very early universe and demonstrated that non-singular bounce models can emerge naturally with quintom matter \cite{Cai:2007qw}. As we know, non-singular bounce cosmology encompasses scenarios in which the Universe initially undergoes a phase of contraction, reaches a finite minimum scale, then subsequently enters an expanding phase. This behavior can be understood by examining the dynamics of the scale factor $a(t)$ and the Hubble parameter $H(t)$. 
It is well-known that in GR, the Friedmann equations that govern the universe are written as:
\begin{equation}
\label{friedmann}
    H^2=\frac{8\pi G}{3}\rho~,~~~\dot H=-4\pi G(\rho+p)~,
\end{equation}
where $H=d\ln a/dt$.
%
When the Universe goes from the contracting phase to the expanding phase, the Hubble parameter goes from $H<0$ to $H>0$, with $H=\rho=0$ at the bounce point. This requires that the time derivative of $H$ must be larger than zero, $\dot H>0$, which gives $\rho+p<0$. Therefore, at this point, one has $w\equiv p/\rho\rightarrow -\infty$, namely below the $w=-1$ line. On the other hand, after the universe enters into the normal expanding phase, it should experience radiation-dominant era ($w=1/3$), matter-dominant era ($w=0$), etc., as required by the standard Big-Bang Theory. This means that in the whole bounce process, the equation of state of our universe will cross $-1$ from below. Such a crossing behavior is the characteristic property of the quintom model. 

After the bounce, the Universe will enter into an expanding phase, and according to its behavior, in the future it will evolve towards various fates \cite{Cai:2009zp}. Among these fates, an interesting one is that the Universe will return to a contracting phase and then bounce again, thus performing a cyclic behavior forever \cite{tolman, Steinhardt:2001st, Steinhardt:2002ih, Brown:2004cs, Piao:2004hr, Piao:2004me, Baum:2006ee}. Like the bounce, a 4D non-singular cyclic universe also needs its EoS to cross $-1$, and thus it can be realized by quintom matter \cite{Feng:2004ff, Xiong:2007cn, Xiong:2008ic}. 

In the following, we will give three models as examples of a quintom bounce, and one example of a cyclic universe with quintom matter.


\subsection{Bounce with two scalar fields}
The simplest Quintom model contains a quintessence-like scalar field and a phantom-like scalar field, with the Lagrangian \cite{Feng:2004ad, Guo:2004fq, Cai:2007qw} \footnote{To be general, the potentials of $\phi$ and $\sigma$ can also include interacting term, see \cite{Zhang:2005eg}.}:
\begin{equation}
\label{twofield quintom}
    S=\int d^4x\sqrt{-g}\left[-\frac{1}{2}\nabla_\mu\phi\nabla^\mu\phi-V(\phi)+\frac{1}{2}\nabla_\mu\sigma\nabla^\mu\sigma-V(\sigma)\right]~,
\end{equation}
where $\phi$ and $\sigma$ stand for quintessence and phantom fields, respectively. The Friedmann equations are:
\begin{eqnarray}
    H^2&=&\frac{8\pi G}{3}\left[\frac{1}{2}\dot\phi^2-\frac{1}{2}\dot\sigma^2+V(\phi)+V(\sigma)\right]~,\\
    \label{bg2}
    \dot H&=&-4\pi G(\dot\phi^2-\dot\sigma^2)~,
\end{eqnarray}
while the equations of motion of the two fields are:
\begin{eqnarray}
    \ddot\phi+3H\dot\phi+V_{,\phi}&=&0~,\\
    \ddot\sigma+3H\dot\sigma+V_{,\sigma}&=&0~.
\end{eqnarray}
From the Friedmann equation one can see that, provided that the potentials of both fields are not less than zero, the only possibility for making $H=0$, as is required by bounce process, is to have the kinetic term of the $\sigma$-field become large and compensate the rest of the energy density. On the other hand, this term should not be dominant in order not to give rise to a negative energy density. This can be achieved by setting $V(\sigma)=0$, so that the equation of motion for the $\sigma$-field becomes $\ddot\sigma+3H\dot\sigma=0$ with the solution $\dot\sigma^2\sim a^{-6}$. Therefore at the bounce point where the scale factor $a(t)$ reaches its minimum, $\dot\sigma^2$ gets its maximum value, while whether the universe goes backwards or forwards, $a(t)$ gets enlarged and $\dot\sigma^2$ will decay. Thus the evolution of the universe away from the bounce depends on the potential of the normal field $\phi$ only. 

To obtain the observational signals, it is important to analyze the perturbations generated by the model. The longitudinal metric perturbations are given by:
\begin{equation}
    ds^2=a^2(\eta)[(1+\Phi)d\eta^2-(1-2\Psi)dx^idx^j]~,
\end{equation}
where $\eta$ is the conformal time with $d\eta=a^{-1}(t)dt$, while $\Phi$ and $\Psi$ are gauge-invariant metric perturbations. The perturbed Einstein equation gives rise to the equations of $\Phi$ and $\Psi$ \cite{Mukhanov:1990me, Piao:2004jg, Cai:2007zv, Cai:2008qb}:
\begin{eqnarray}
    \nabla^2\Psi-3{\cal H}(\Psi'+{\cal H}\Phi)&=&4\pi G[\phi'(\delta\phi'-\phi'\Phi)+a^2V_{,\phi}\delta\phi-\sigma'(\delta\sigma'-\sigma'\Phi)]~,\\
    \Psi'+{\cal H}\Phi&=&-4\pi G [-\phi'\delta\phi+\sigma'\delta\sigma]~,\\
    \Phi''+3{\cal H}\Phi'+(2{\cal H}'+{\cal H}^2)\Phi&=&4\pi G [\phi'(\delta\phi'-\phi'\Phi)-a^2V_{,\phi}\delta\phi-\sigma'(\delta\sigma'-\sigma'\Phi)]~,
\end{eqnarray}
which can be combined to give the main equation of the Newtonian potential $\Phi$:
\begin{equation}
\label{perteq}
    \Phi''+2\left({\cal H}-\frac{\phi''}{\phi'}\right)\Phi'+2\left({\cal H}'-{\cal H}\frac{\phi''}{\phi'}\right)\Phi-\nabla^2\Phi=8\pi G\left(2{\cal H}+\frac{\phi''}{\phi'}\right)\sigma'\delta\sigma~,
\end{equation}
where prime denotes derivative with respect to the conformal time $\eta$, and ${\cal H}$ is the conformal Hubble parameter, ${\cal H}\equiv a'/a$. Moreover, the curvature fluctuation $\zeta$ is related with $\Phi$ via:
\begin{equation}
    \zeta=\Phi+\frac{\cal H}{{\cal H}^2-{\cal H}'}(\Phi'+{\cal H}\Phi)~,
\end{equation}
which is conserved on super-Hubble scales:
\begin{equation}
    (1+w)\zeta'=\frac{2k^2(\Phi'+{\cal H}\Phi)}{9{\cal H}^2}~.
\end{equation}

In the following, we will see that different forms of potential $V(\phi)$ will give rise to different backgrounds as well as different evolutions of the perturbations. In order to understand their main influence, we consider two broad categories of potentials, namely large field potentials and small field potentials.
 
\subsubsection{Large field potentials}
 
The typical large field potential is the mass squared potential, $V(\phi)=m^2\phi^2/2$ \cite{Cai:2007zv}. This potential is symmetric with respect to its minimum at $\phi=0$, therefore it can give rise to symmetric evolution before and after the bounce. 

\begin{figure}[htbp]
    \centering
    \begin{minipage}[b]{0.48\linewidth}
        \includegraphics[width=1\linewidth]{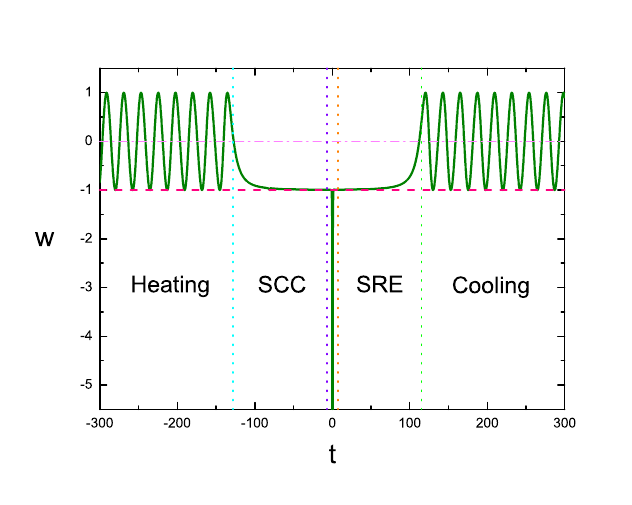}
    \caption{Plot of evolution of EoS $w$ in a double-field quintom bounce with large field potential. The figure is taken from Ref.~\cite{Cai:2007zv}. }
    \label{eos_twofield1}
    \end{minipage}
\hfill
    \begin{minipage}[b]{0.48\linewidth}
        \includegraphics[width=0.7\linewidth]{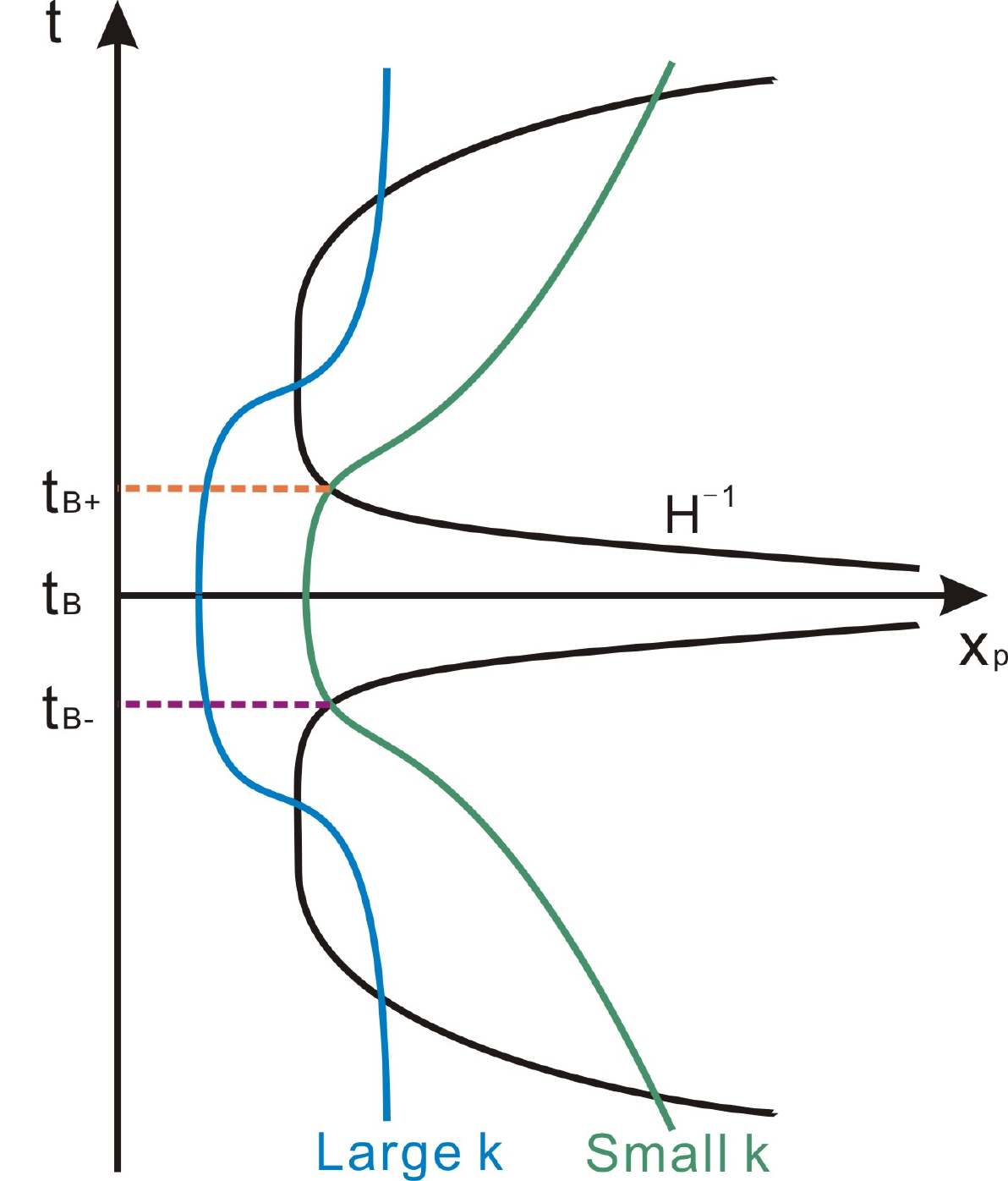}
    \caption{The evolution of perturbation wavelengths with different comoving wavenumbers $k$ as well as the Hubble radius $|H^{-1}|$. The figure is taken from Ref.~\cite{Cai:2007zv}. }
    \label{largefieldplot}
    \end{minipage}
\end{figure}

At the very beginning, the $\phi$-field is set near the minimum and oscillates with increasing amplitude, due to the contraction of the Universe and the friction term. This will cause an oscillating behavior of the EoS, with the average value of $\langle w\rangle=0$. This phase is called the ``heating phase". When the friction becomes less important, the Universe enters into a ``slow-climbing" phase, where $\phi$ evolves slowly along the potential upwards, with the EoS $w\simeq-1$. Meanwhile, the kinetic term of the $\sigma$-field becomes more and more important. When it reaches the value of the $\phi$-field's energy density, the two parts cancel out and the bounce takes place. At the bounce point, the EoS reaches $-\infty$, as has been demonstrated previously. After the bounce, the $\sigma$-field decays again, and the $\phi$-field also rolls slowly along the potential to its minimum, and the EoS approaches $-1$ again. Finally, when the $\phi$-field reaches the bottom, it oscillates around its minimum again with decreasing amplitude, with $\langle w\rangle=0$. The evolution behavior of $w$ through the whole process is shown in Fig. \ref{eos_twofield1}.

The evolution of the perturbation is given by the main equation \eqref{perteq}. Before going into the details, we draw the evolution of perturbation wavelength scales vs. Hubble radius in Fig. \ref{largefieldplot}. From this plot we can see that, the evolution is symmetric before and after the bounce, thus like the expanding phase, in the contracting phase, the Hubble radius will also shrink and expand (to infinity at the bounce point), and the perturbations will also exit and reenter the horizon. This will change the evolution of the perturbations and will affect the final observational signals in the CMB. Moreover, for large $k$ (small scale) modes, the reentrance into the horizon takes place in the slow-climbing region, for small $k$ (large scale) modes it happens in the bounce phase. Therefore the evolution of the perturbations is highly nontrivial. 

In the heating phase, since the EoS is effectively zero, one has
\begin{equation}
\label{heatingbg}
    a\propto \eta^2~,~~~{\cal H}=\frac{2}{\eta}~,~~~\phi'=\frac{1}{\eta}~.
\end{equation}
Moreover, since in this phase the kinetic term of $\sigma$-field is negligible, we can omit the right hand side of Eq. \eqref{perteq} for simplicity. Then the equation becomes (in momentum space)
\begin{equation}
\label{heatingpert}
    \Phi''_k+\frac{6}{\eta}\Phi'_k+k^2\Phi_k=0~,
\end{equation}
with the solution 
\begin{equation}
\label{heatingsol}
    \Phi_k=\eta^{-5/2}[k^{-5/2}A_kJ_{5/2}(k\eta)+k^{5/2}B_kJ_{-5/2}(k\eta)]~.
\end{equation}
The coefficients can be determined by matching the sub-horizon approximation ($|k\eta|\gg 1$) with the initial condition of $\Phi_k$, which we take to be the Bunch-Davies vacuum solution:
\begin{equation}
    \Phi_k=A\eta^{-3}\frac{e^{-ik\eta}}{\sqrt{2k^3}}~,~~~A\equiv4\pi G\sqrt{\rho_0}\eta_0^3~,
\end{equation}
from which we get $A_k=(i\sqrt{\pi}/2)Ak^{3/2}$, $B_k=-(\sqrt{\pi}/2)Ak^{-7/2}$. On the other hand, in the super-horizon approximation with $|k\eta|\ll 1$, the approximate solution becomes:
\begin{equation}
\label{heatingsolsup}
    \Phi_k\simeq \frac{iAk^{3/2}}{15\sqrt{2}}-\frac{3A}{\sqrt{2}}k^{-7/2}\eta^{-5}~,
\end{equation}
which has one constant mode and one growing mode.

We assume that the Universe enters into the slow-climbing phase at the time $\eta_c$. After that, Eq. \eqref{perteq} becomes
\begin{equation}
\label{climbpert}
    \Phi''_k+(2\epsilon_H-\delta_H){\cal H}\Phi'_k-\delta_H{\cal H}^2\Phi_k+k^2\Phi_k=0~,
\end{equation}
where we defined the slow-climb parameters: $\epsilon_H\equiv-\dot H/H^2$, $\delta_H\equiv\dot\epsilon_H/(H\epsilon_H)$, satisfying $|\epsilon_H|,|\delta_H|\ll 1$. The solution is
\begin{equation}
\label{slowclimbsol}
    \Phi_k=(\eta-\tilde\eta_c)^\alpha[k^{-\nu}C_kJ_\nu(k(\eta-\tilde\eta_c))+k^\nu D_kJ_{-\nu}(k(\eta-\tilde\eta_c))]~,
\end{equation}
where $\alpha\simeq 1/2$, $\nu\simeq 1/2$, and $\tilde\eta_c\equiv \eta_c+1/{\cal H}_c$. The super-horizon approximation, which is to be connected with the super-horizon approximation in the heating phase \eqref{heatingsolsup}, turns out to be:
\begin{equation}
\label{slowclimbsolsup}
    \Phi_k\simeq \sqrt{\frac{2}{\pi}}[C_k(\eta-\tilde\eta_c)+D_k]~.
\end{equation}

By matching with \eqref{heatingsolsup} using matching conditions \cite{Hwang:1991an, Deruelle:1995kd} (see also \cite{Durrer:2002jn}), one gets the coefficients:
\begin{eqnarray}
    C_k&=&-{\cal H}_c\left[\frac{1}{15}\left(1-\frac{2}{3}\epsilon_H\right)A_k+3(1+\epsilon_H)B_k\eta_c^{-5}\right]~,\\
    D_k&=&\epsilon_H\left(\frac{2}{45}A_k-3B_k\eta_c^{-5}\right)\simeq 0~.
\end{eqnarray}
Thus the main contribution comes from the growing mode. On the other hand, the sub-horizon approximation of \eqref{slowclimbsol} is
\begin{equation}
\label{slowclimbsolsub}
    \Phi_k\simeq\sqrt{\frac{2}{\pi}}\left[\frac{C_k}{k}\sin(k(\eta-\tilde\eta_c))+D_k\cos(k(\eta-\tilde\eta_c))\right]~.
\end{equation}

When the universe enters into a bounce phase, the EoS goes down towards $-\infty$ and goes up again to above $-1$, so it will be complicated to analyze the dynamics of the perturbations. Nevertheless, it is convenient to parametrize the Hubble parameter as a linear function which crosses zero at the bounce point, namely
\begin{equation}
    {\cal H}\simeq \frac{y}{2} (\eta-\eta_B)~,~~~a\simeq \frac{a_B}{1-y(\eta-\eta_B)^2/4}~,
\end{equation}
where $\eta_B$ is the time when bounce occurs, and $a_B$ is the scale factor at $\eta_B$. Moreover, during the bounce phase $\dot\sigma$ becomes important.  However, from the background equation \eqref{bg2} one approximately has $\phi''\simeq-2{\cal H}\phi'$, so the right hand side of Eq. \eqref{perteq} can still be neglected. Therefore, the perturbation equation during the bounce phase becomes:
\begin{equation}
\label{bouncepert}
    \Phi''_k+3y(\eta-\eta_B)\Phi'_k+(k^2+y)\Phi_k\simeq 0~,
\end{equation}
and the solution is
\begin{equation}
\label{bouncesol}
    \Phi_k\simeq e^{-\frac{3}{4}y(\eta-\eta_B)^2}\left[E_k\sin[k(\eta-\eta_B)]+F_k\cos[k(\eta-\eta_B)]\right]~.
\end{equation}

The coefficients in solution \eqref{bouncesol}, namely $E_k$ and $F_k$, are obtained by matching \eqref{bouncesol} to the slow-climbing solution \eqref{slowclimbsol} with the same matching conditions. From Fig. \ref{largefieldplot}, one can see that there are two cases, namely matching in the sub-horizon region (for large $k$) and matching in the super-horizon region (for small $k$). In the first case \eqref{bouncesol} matches \eqref{slowclimbsolsub}, while in the second case it matches \eqref{slowclimbsolsup}. This will give two sets of $E_k$ and $F_k$, and the detailed expressions are given in \cite{Cai:2007zv}. From the solution we can also see that, while in the first case $\Phi_k$ in the bounce phase is dominated by the growing mode in slow-climbing phase, in the second case it is dominated by the constant mode. 

After the bounce, the Universe enters into an expanding phase at the time $\eta_{B+}$, where the $\phi$ field is slow-rolling. In this phase, the perturbation equation is the same as Eq. \eqref{climbpert}, with the solution:
\begin{equation}
\label{slowrollsol}
    \Phi_k=(\eta-\tilde\eta_{B+})^\alpha[k^{-\nu}G_kJ_\nu(k(\eta-\tilde\eta_{B+}))+k^\nu H_kJ_{-\nu}(k(\eta-\tilde\eta_{B+}))]~,
\end{equation}
where $\tilde\eta_{B+}\equiv\eta_{B+}+1/{\cal H}_{B+}$. The sub-horizon approximation is:
\begin{equation}
\label{slowrollsolsub}
    \Phi_k\simeq\sqrt{\frac{2}{\pi}}\left[\frac{G_k}{k}\sin(k(\eta-\tilde\eta_{B+}))+H_k\cos(k(\eta-\tilde\eta_{B+}))\right]~,
\end{equation}
while the super-horizon approximation is:
\begin{equation}
\label{slowrollsolsup}
    \Phi_k\simeq \sqrt{\frac{2}{\pi}}[G_k(\eta-\tilde\eta_{B+})+H_k]~,
\end{equation}
where the first and second terms correspond to the decaying and constant mode, respectively. The coefficients $G_k$ and $H_k$ will also be obtained by matching the solution \eqref{slowrollsol} to the solution \eqref{bouncesol} in the bounce phase. Similarly, there are two cases of matching at sub-horizon and super horizon regions, where \eqref{bouncesol} matches to \eqref{slowrollsolsub} and \eqref{slowrollsolsup}, respectively. The detailed expressions are given in \cite{Cai:2007zv}. In the first case where both $G_k$ and $H_k$ are important, the final perturbation $\Phi_k$ is dominated by its constant mode. While in the second case where $H_k\simeq 0$ is obtained, $\Phi_k$ is dominated by its decaying mode \cite{Veneziano:1991ek, Gasperini:1992em, Gasperini:2002bn, Khoury:2001wf, Khoury:2001bz, Steinhardt:2001st, Brustein:1994kn, Lyth:2001pf, Lyth:2001nv, Finelli:2001sr, Hwang:2001ga, Creminelli:2004jg, Tolley:2003nx, Battefeld:2005wv}.

CMB observations suggest a nearly scale-invariant power spectrum of primordial curvature perturbations, which could be realized by either de-Sitter like expansion (inflation) or matter-like contraction. However, due to the addition of the slow-climbing phase, the $k$-dependence of the perturbations become more complicated, and in general cannot give rise to a scale-invariant power spectrum (see numerical results in \cite{Cai:2007zv}). 

\subsubsection{Small field potential}

The typical small field potential is the Coleman-Weinberg potential, $V(\phi)=(\lambda\phi^4/4)\ln(|\phi|/v)-\lambda(\phi^4- v^4)/16$ \cite{Cai:2008qb, Coleman:1973jx}. As a result of this potential, the vacuum is shifted to $\phi=\pm v$.  Therefore, the symmetry is spontaneously broken in the vacuum. It will also give rise to an asymmetric bounce behavior. 

\begin{figure}[htbp]
    \centering
    \begin{minipage}[b]{0.48\linewidth}
        \includegraphics[width=1.1\linewidth]{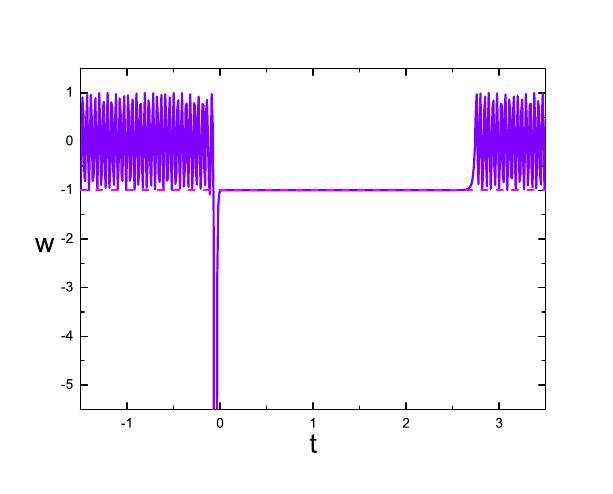}
    \caption{Plot of evolution of EoS $w$ in a double-field quintom bounce with small field potential. The figure is taken from Ref.~\cite{Cai:2008qb}. }
    \label{eos_twofield2}
    \end{minipage}
\hfill
    \begin{minipage}[b]{0.48\linewidth}
        \includegraphics[width=0.6\linewidth]{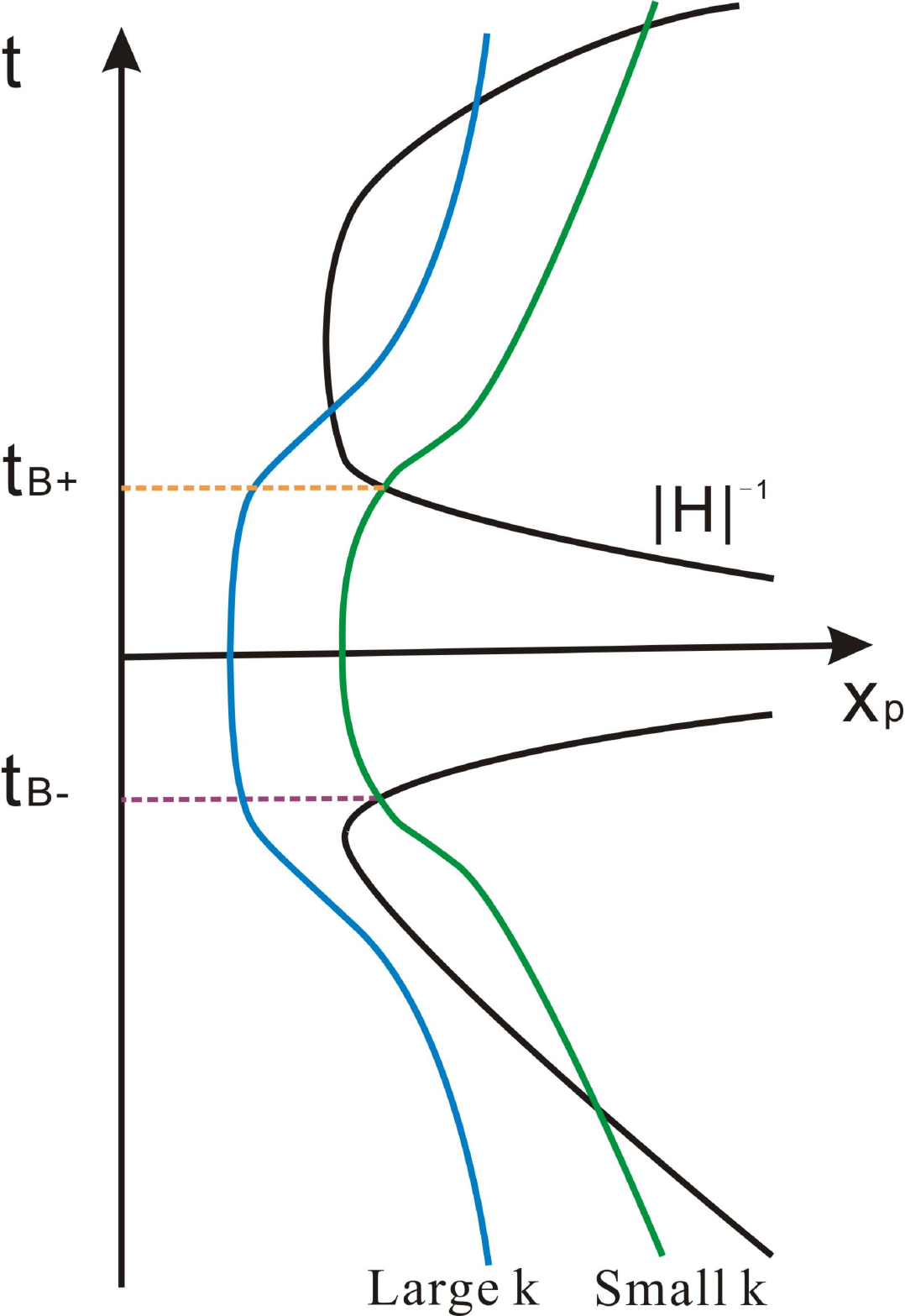}
    \caption{The evolution of perturbation wavelengths with different comoving wavenumbers $k$ as well as Hubble radius $|H^{-1}|$. The figure is taken from Ref.~\cite{Cai:2008qb}.}
    \label{smallfieldplot}
    \end{minipage}
\end{figure}

Fig. \ref{eos_twofield2} shows the evolution behavior of $w$ over the process in this model. Initially, the $\phi$ field is still set around the minimum value, and it can oscillate with its amplitude increasing and $\langle w\rangle=0$. When the field reaches the plateau, the $\sigma$ field catches up, and the bounce takes place when $w$ goes to minus infinity. After the bounce, when the $\phi$-field rolls along the plateau, the Universe enters into a slow-roll phase where $w\simeq-1$, very much like inflation. Finally when the field drops into another minimum and oscillates, the inflation ends and the field becomes oscillating again, with $\langle w\rangle=0$.

We also draw the evolution plot of perturbation wavelength scales v.s. Hubble radius in Fig. \ref{smallfieldplot}. From the plot we can see that, since now the symmetry before and after the bounce is broken, the evolution of the perturbations will also be different. While the small $k$ modes of perturbations still need to exit and reenter the horizon in the contracting phase, the large $k$ modes needn't, and will stay inside the horizon until the expanding phase. In the following, we will focus on these modes.

In the contracting phase where the EoS of the Universe oscillates continuously with $\langle w\rangle=0$, the background and perturbation evolutions are still described by Eqs. \eqref{heatingbg} and \eqref{heatingpert}. While the solution is given in Eq. \eqref{heatingsol}, since we are only interested in the subhorizon solution, we only have:
\begin{equation}
\label{BDvacuum}
    \Phi_k=4\pi G\frac{\sqrt{\rho_0}\eta_0^3}{\eta^3}\frac{e^{-ik\eta}}{\sqrt{2k^3}}~.
\end{equation}

Since in this case the evolution is not symmetric and for large $k$ modes there is no horizon-exit behavior before bounce, we don't have a ``slow-climbing" phase. In bounce phase, the equation is the same as \eqref{bouncepert}, and the solution is given by \eqref{bouncesol}. After the bounce, the universe will enter a slow-rolling expanding phase, in which the perturbation equation is \eqref{climbpert} and the solution is the same as \eqref{slowrollsol}. 

The coefficients $E_k$, $F_k$, $G_k$ and $H_k$ are obtained using the same matching method as above. However, since here the matchings are only performed at sub-horizon regions, the final perturbations are determined by the constant mode of $H_k$. From the tedious calculation done in \cite{Cai:2008qb}, one obtains 
\begin{equation}
    H_k=\sqrt{\frac{\pi}{2}}\frac{4\pi G}{\sqrt{2k^3}}|\dot\phi|e^{-ik\tilde\eta_{B+}}\left\{1+\frac{3e^{-ik(\eta_{B-}-\tilde\eta_{B+})}}{k\eta_{B-}}\sin[k(\eta_{B-}-\tilde\eta_{B+})]\right\}~.
\end{equation}
where $\eta_{B-}$ is the time when the Universe enters the bounce phase. Moreover, in the super-horizon region with nearly de-Sitter spacetime we have $\zeta\simeq \Phi/\epsilon_H$, Thus one obtains the power spectrum of the curvature perturbations
\begin{equation}
\label{smallfieldspec}
    P_\zeta\equiv\frac{k^3}{2\pi^2}|\zeta|^2=\frac{8G^2\rho}{3\epsilon_H}\left\{1-\frac{3{\cal H}_{B-}}{2k}\sin\frac{2k}{{\cal H}_{B+}}\right\}~.
\end{equation}
This result indicates that for large $k$ (small scale) region where the last term could be neglected, the power spectrum is both constant and scale-invariant, while for small $k$ (large scale) region where the second term becomes important, the spectrum gets some wiggle-like corrections. This is due to the effect of perturbations during bounce region. Moreover, for even larger scales, as the perturbations exit the horizon before the bounce, the power spectrum will be blue-tilted and will thus get suppressed \cite{Piao:2003zm, Piao:2005ag, Liu:2013kea}. There are already some hints of these signals in the observations \cite{Planck:2018vyg, Planck:2018jri, Planck:2019evm}. If they are furtherly confirmed, they will provide a smoking gun for a bouncing cosmology.   

\subsection{Bounce with a higher-derivative field}
The single field Quintom model with higher-derivative interactions was proposed in \cite{Li:2005fm}. With higher-derivative terms, the single field gains an additional degree of freedom that can make the EoS cross $-1$. When applied to the early Universe, it can also give rise to a bounce solution.

\begin{figure}[htbp]
    \centering
    \begin{minipage}[b]{0.48\linewidth}
        \includegraphics[width=1\linewidth]{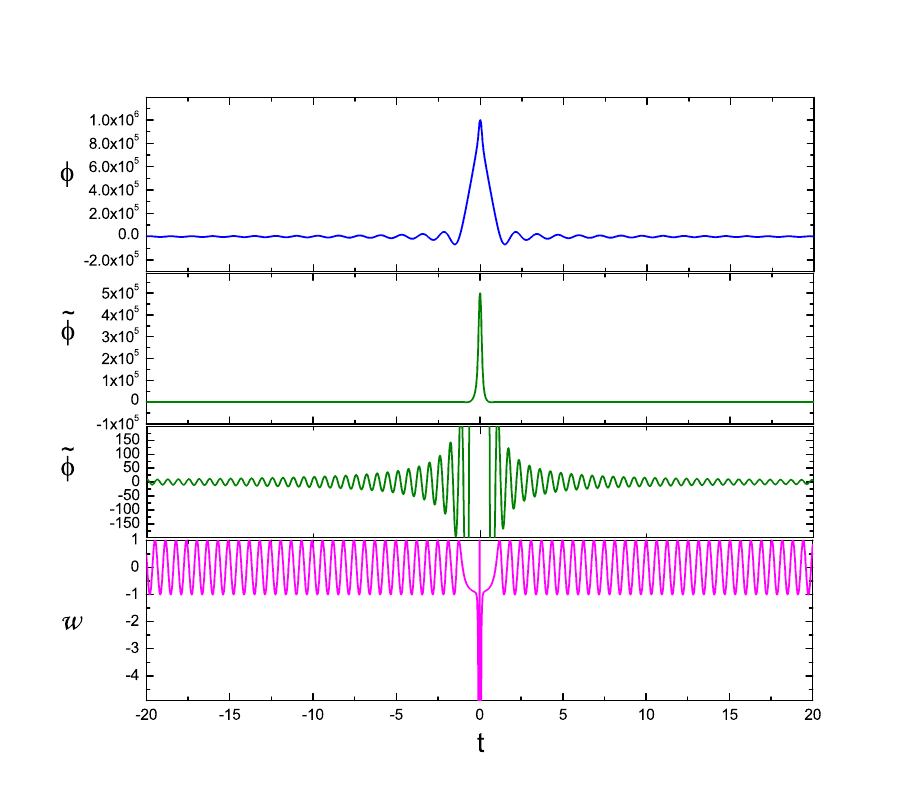}
    \caption{Plot of evolutions of fields $\phi$, $\tilde\phi$ and EoS $w$ in the Lee-Wick type quintom bounce. The figure is taken from Ref.~\cite{Cai:2008qw}.}
    \label{LWeos}
    \end{minipage}
\hfill
    \begin{minipage}[b]{0.48\linewidth}
        \includegraphics[width=0.7\linewidth]{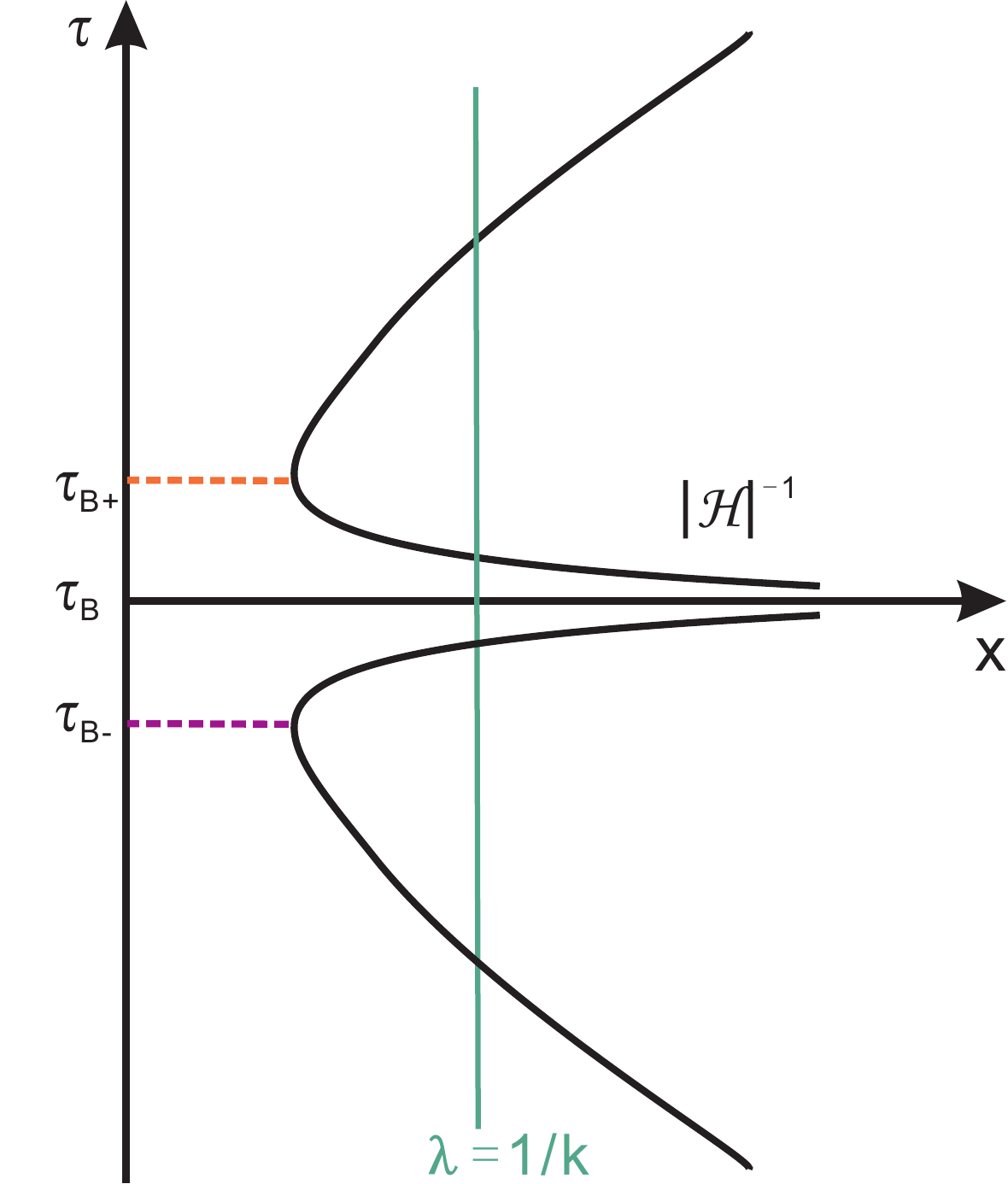}
    \caption{The comoving wavelength $\lambda =1/k$ as well as the comoving Hubble radius $|{\cal H}^{-1}|$. The figure is taken from Ref.~\cite{Cai:2008qw}.}
    \label{LWplot}
    \end{minipage}
\end{figure}

A detailed analysis of a bounce model with higher-derivative interactions was performed in Ref.~\cite{Cai:2008qw}. In this work, the Lagrangian of the single field $\hat{\phi}$ is written as:
\begin{equation}
\label{hd quintom}
    L=\frac{1}{2}\nabla_\mu\hat\phi\nabla^\mu\hat\phi-\frac{1}{2M^2}(\Box\hat\phi)^2-\frac{1}{2}m^2\hat\phi^2~.
\end{equation}
In particle physics, this Lagrangian is called the ``Lee-Wick model", in which the Higgs mass is stabilized against radiative corrections to solve the hierarchy problem \cite{Lee:1969fy, Lee:1970iw, Grinstein:2007mp}. Note that this model can also be transformed into a two-scalar-field model. By introducing an auxiliary field $\tilde\phi$ and redefining a scalar field as:
\begin{equation}
    \phi=\hat\phi+\tilde\phi~,
\end{equation}
the Lagrangian \eqref{hd quintom} then becomes
\begin{equation}
    L=\frac{1}{2}\nabla_\mu\phi\nabla^\mu\phi-\frac{1}{2}\nabla_\mu\tilde\phi\nabla^\mu\tilde\phi-\frac{1}{2}m^2\phi^2+\frac{1}{2}M^2\tilde\phi^2~,
\end{equation}
where the equations of motion (EoM) for the $\phi$ and $\tilde\phi$ fields are 
\begin{equation}
    \ddot\phi+3H\dot\phi+m^2\phi=0~,~~~\ddot{\tilde\phi}+3H\dot{\tilde\phi}+M^2\tilde\phi=0~,
\end{equation}
respectively. From the EoM one can see that, since both fields now get masses, in the contracting phase they will both oscillate around their vacuum with increasing amplitude,
\begin{equation}
    |\phi(t)|\sim |\tilde\phi(t)|\sim a^{-3/2}(t)~,
\end{equation}
and will freeze out when their amplitude reaches $(12\pi)^{-1/2}m_{pl}$. At the initial time let's assume the energy density of $\phi(t)$ is larger than $\tilde\phi(t)$, so that $\phi(t)$ will dominate the Universe. When $\phi(t)$ is oscillating, the Universe will behave like matter with its average value of EoS $\langle w\rangle=0$. Similarly to the large field bounce story, when the $\phi$-field reaches $(12\pi)^{-1/2}m_{pl}$ and then freezes out, it will slow-climb along the potential and tend to drive a period of ``deflation". But what is different is that, now the second field $\tilde\phi$ also has mass and will also oscillate, which will destroy the deflation behavior. If the $\tilde\phi$ field oscillates until its energy density catches up with that of $\phi$, the total energy density will vanish and the bounce will occur. If the bounce happens before $\tilde\phi$ freezes out, there will be no deflation era and we will get a pure matter bounce \cite{Finelli:2001sr, Wands:1998yp, Peter:2008qz}. Moreover, since both fields have symmetric potentials, the evolution of the Universe will also be symmetric before and after the bounce, namely the Universe will directly enter into a matter-dominated era without inflation. The evolutions of both fields and the EoS of the Universe are plotted in Fig. \ref{LWeos}.

The perturbation equation is the same as in case of the two-field models, namely Eq. \eqref{perteq}.  However, as is described above, the cosmological periods that the Universe experiences are different. Here in Fig. \ref{LWplot} we plot the perturbation wavelength scales vs. Hubble radius (in conformal time). The process is just from matter dominance to the bounce, then to matter dominance again. As is demonstrated above, the initial condition is given by the Bunch-Davies vacuum solution, namely Eq. \eqref{BDvacuum}. 

In the previous section, we have shown that if the matter contracting phase is followed by an inflationary phase after the bounce, where the perturbations did not exit the horizon before the bounce, a scale-invariant power spectrum is acquired. This is because the evolution inside of the horizon actually does not affect the averaged $k$-dependence of the perturbations and it still behaves like the Bunch-Davies vacuum. However, in this model, in the absence of inflation era, one may not be able to obtain a scale-invariant power spectrum. As we will see below, this issue can be overcome by making the perturbations exit the horizon before the bounce. In this case, the growing mode of $\Phi$, which dominates in the contracting phase, gets deeply red-tilted, and when crossing the bounce point, it will be compensated by the same amount of blue-tilting.  

The super-horizon approximation of the contracting phase was obtained previously
\begin{equation}
\label{LWcontractsol}
    \Phi=D_-+S_-(\eta-\tilde\eta_{B-})^{-5}~,
\end{equation}
where $D_-\sim k^{3/2}$, $S_-\sim k^{-7/2}$, $\tilde\eta_{B-}\equiv \eta_{B-}-2/{\cal H}_{B-}$. While the first mode is constant, the second mode is growing.

During the bounce phase, the equation of motion is given by Eq. \eqref{bouncepert}. However, since we are considering the fluctuation modes which exit the horizon in the heating phase, unlike the case of the previous solution \eqref{bouncesol}, now we consider the solution with much smaller $k$, namely $k^2\ll y$. Thus the $y$-term will dominate $k^2$ term and the solution will present approximately no $k$-dependence. This is a very important property of the large-scale solution which will affect the scale-dependence as follows. The solution is given by
\begin{equation}
\label{LWbouncesol}
    \Phi_k=F_k+E_k\sqrt{y}(\eta-\eta_B)-\frac{3}{5}F_ky(\eta-\eta_B)^2+{\cal O}(y^{3/2}(\eta-\eta_B)^3)~,
\end{equation}
where we assume that the bounce period is very short and it is enough to expand up to ${\cal O}(\eta-\eta_B)^3$.

\begin{figure}[htbp]
    \centering
        \includegraphics[width=0.45\textwidth]{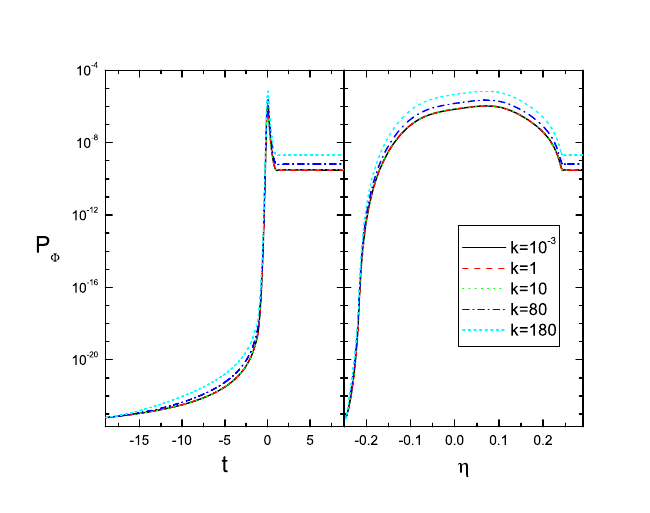}
        \includegraphics[width=0.5\textwidth]{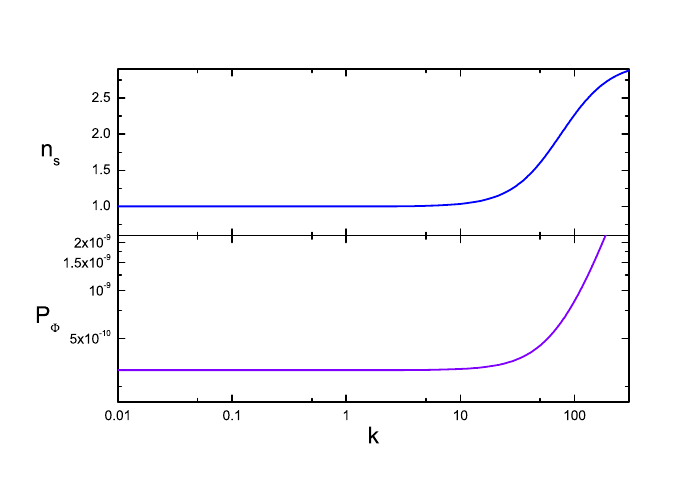}
    \caption{Left: numerical results for the power spectrum of curvature perturbations $P_\Phi$ with respect to time ($t$ for cosmic time and $\eta$ for comoving time). Right: the power spectrum $P_\Phi$ as well as the spectral index $n_s$ with respect to comoving wavenumber $k$. The figures are taken from Ref.~\cite{Cai:2008qw}.}
    \label{LWspectrum}
\end{figure}

In the expanding phase, the solution is similar to the contracting phase since both have an EoS of matter
\begin{equation}
\label{LWexpandsol}
    \Phi=D_++S_+(\eta-\tilde\eta_{B+})^{-5}~,
\end{equation}
where $\tilde\eta_{B+}\equiv \eta_{B+}-2/{\cal H}_{B+}$. Now the second mode becomes decaying, so the constant mode will be dominant. Matching the solutions \eqref{LWcontractsol}, \eqref{LWbouncesol} and \eqref{LWexpandsol} one obtains the final result for $D_+$:
\begin{eqnarray}
\label{D+sol}
    D_+&=&D_-+\left[-\frac{4}{5}D_-+\frac{3}{5}\frac{{\cal H}_{B-}^5}{2^4}S_-\right]\frac{2k^2}{9{\cal H}_{B-}^2}~,\nonumber\\
    &=&-\frac{\sqrt{\rho_{B-}}}{10\sqrt{2}}k^{-3/2}~,
\end{eqnarray}
where in the last line we made use of $D_-\sim k^{3/2}$, $S_-\sim k^{-7/2}$. Here $\rho_{B-}$ is the energy density at $\eta_{B-}$. Therefore the power spectrum turns out to be
\begin{equation}
    P_\Phi\equiv\frac{k^3}{2\pi^2}|D_+|^2=\frac{\rho_{B-}}{(20\pi)^2}~,
\end{equation}
which is scale-invariant. From Eq. \eqref{D+sol} one can see that, this is due to the fact that in the small $k$ region, while the constant mode is suppressed, the growing mode, which is deep red, will contribute to $D+$, and will be blue-tilted by the right amount. In Fig. \ref{LWspectrum} we plot $P_\Phi$ and the spectral index $n_s$ with respect to $k$. We can see from the plot that, while in the large $k$ region the power spectrum is blue-tilted, in the small $k$ region it will be scale-invariant.  

For the matter bounce, initial fluctuations could also be generated classically by the thermal matter perturbations inside the Hubble radius \cite{Nayeri:2005ck, Brandenberger:2006pr, Brandenberger:2006vv, Cai:2008qw, Qiu:2010vs}. From the perturbed Einstein equations, the Fourier space correlation function of the Newtonian potential $\Phi$ is related to density fluctuations $\langle|\delta T^0_0(k)|\rangle^2$ (or mass fluctuations in position space $\delta M(R)$),
\begin{equation}
    \langle|\Phi(k)|\rangle^2=16\pi^2G^2k^{-4}a^4\langle|\delta T^0_0(k)|\rangle^2=16\pi^2G^2k^{-4}a^4R^{-3}\delta M(R)^2~,
\end{equation}
where $R=a/k$ is the physical radius of the region corresponding to the wavenumber $k$. On the other hand, $\delta M(R)^2\sim T^5 R^3$.  Therefore one has
\begin{equation}
    P_\Phi(k)\sim k^3\langle|\Phi(k)|\rangle^2\sim k^{-1}T^5\sim k^{-1}a^{-5}~,
\end{equation}
where we used the fact $T\sim a^{-1}$ for thermal matter. At horizon crossing time $t_c$ where $a(t_c)H(t_c)\sim k$, one can get $a\sim k^{-(1+3w)/2}$. For matter contraction $\langle|w|\rangle=0$, this leads to $P_\Phi\sim k^9$ (constant mode) and $k^{-1}$ (growing mode) \cite{Cai:2008qw}. 

One can also discuss the tensor perturbations in the bounce model, which contribute to the primordial gravitational waves and will have signals in the B-mode polarizations of the CMB map. Since the tensor perturbation is also generated in the contracting phase and propagates during the contracting and bounce phases, it is expected to be different from that generated in the normal inflationary scenario. To see this, one can start with the FRW metric with tensorial perturbation
\begin{equation}
    ds^2=a^2(\eta)[d\eta^2-(\delta_{ij}+h_{ij})dx^idx^j]~,
\end{equation}
with the properties $h_{ij}=h_{ji}$, $h_{ii}=0$ and $h_{ij,j}=0$. It is well known that, in Einstein's gravity, it satisfies the equation of motion
\begin{equation}
    h_{ij}''+2\frac{a'}{a}h_{ij}'-\nabla^2h_{ij}=0~.
\end{equation}
Considering the Fourier modes of $h_{ij}$
\begin{equation}
    h_{ij}=\sum_{\lambda=1,2}\int \frac{d^3k}{(2\pi)^{3/2}}h^{(\lambda)}_k(\eta,{\bf k})e^{(\lambda)}_{ij}e^{i{\bf k}{\bf x}}~,
\end{equation}
and defining $v_k=ah_k/\sqrt{2}$ (omitting superscript ``$(\lambda)$"), one gets the Mukhanov-Sasaki equation for $v_k$:
\begin{equation}
\label{perteqtensor}
    v_k''+\left(k^2-\frac{a''}{a}\right)v_k=0~.
\end{equation}

The procedure for dealing with this equation is the same as that for scalar perturbations. The solution is divided into three stages, namely the contracting and expanding phases with $w=0$, and the bounce phase. For the former, the behavior of $a(\eta)$ is given by Eq. \eqref{heatingbg}, so one has $a''/a\propto \eta^{-2}$. For the latter one has
\begin{equation}
    \frac{a''}{a}\simeq \frac{4}{\pi}\alpha a_B^2=\frac{y}{3}~.
\end{equation}
One can then solve Eq. \eqref{perteqtensor} for each stage and connect with matching conditions. The final solution for $v_k$ will take an asymptotic form,
\begin{equation}
    v_k\simeq i\frac{\sqrt{2}}{24}\frac{{\cal H}^3_{B+}}{k^{3/2}}(\eta-\tilde\eta_{B+})^2~.
\end{equation}
Then the tensor power spectrum is given by
\begin{equation}
    P_T(k)\equiv 2\frac{k^3}{\pi^2}\left|\frac{v_k}{a}\right|^2=\frac{2\rho_{B+}}{27\pi^2}~,
\end{equation}
indicating a scale-invariant and constant tensor power spectrum \cite{Cai:2008qw, Cai:2008ed}. 

In 2008, a class of single scalar models called the Galileon has been proposed \cite{Nicolis:2008in, Deffayet:2009wt, Nicolis:2009qm}. These models contain higher derivative terms in their Lagrangians, however, due to the delicate design of the Lagrangians (for example, the coupling of the higher derivative term to the kinetic term), their equations of motion can be made 2nd order, therefore avoiding a dynamical ghost mode. These models can be extended to include modified gravity or nonminimal coupling, which is a realization of an idea pioneered by Horndeski in 1974 \cite{Horndeski:1974wa, Kobayashi:2019hrl}. Due to this interesting property, these models have been applied to many aspects in cosmology, see Ref.~\cite{Chow:2009fm, Gannouji:2010au, Kobayashi:2010cm, Deffayet:2010qz, Burrage:2010cu, VanAcoleyen:2011mj, Liu:2011ns, Deffayet:2011gz, Evslin:2011vh, DeFelice:2011uc}.

\begin{figure}[htbp]
    \centering
        \includegraphics[width=0.6\textwidth]{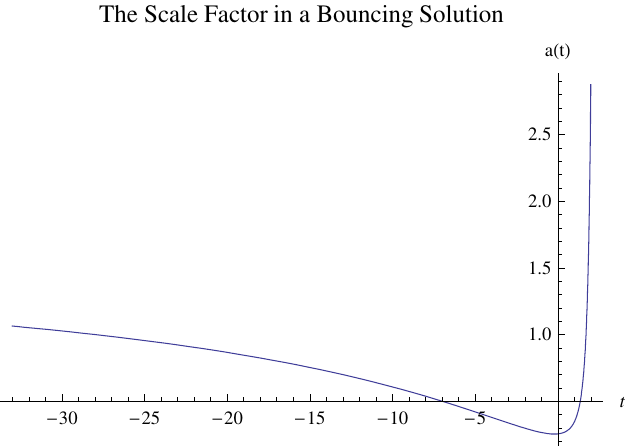}
    \caption{The evolution of the scale factor $a$ with respect to cosmic time $t$ in a Galileon type quintom bounce. The figure is taken from Ref.~\cite{Qiu:2011cy}.}
    \label{Galileona}
\end{figure}

A Galileon model can also realize quintom behavior and drive a bounce \cite{Qiu:2011cy, Easson:2011zy, Cai:2012va, Qiu:2013eoa, Qiu:2014nla, Qiu:2015nha, Wan:2015hya, Cai:2016thi, Cai:2017tku, Ni:2017jxw, Qiu:2024sdd, Li:2024rgq}. Here we give a concrete model:
\begin{equation}
\label{actionG}
    S=\int d^4x\sqrt{-g}\left[\frac{R}{16\pi G}+F^2 e^{2\phi}(\partial\phi)^2+\frac{F^3}{M^3}(\partial\phi)^2\Box\phi+\frac{F^3}{2M^3}(\partial\phi)^4\right]~,
\end{equation}
which is called the conformal Galileon model \cite{Creminelli:2010ba}. The coefficient $F$ is the mass scale of the Galileon field, while $M$ is the scale of higher derivative operators. Thus we can get the energy density and pressure as
\begin{eqnarray}
    \rho&=&F^2\left[-e^{2\phi}\dot\phi^2+\left(\frac{3F}{2M^3}\right)(\dot\phi^4+4H\dot\phi^3)\right]~,\\
    p&=&F^2\left[-e^{2\phi}\dot\phi^2+\left(\frac{F}{2M^3}\right)(\dot\phi^4-4\dot\phi^2\ddot\phi)\right]~.
\end{eqnarray}
In the absence of potential terms in $\rho$ and $p$, the Friedmann equation can be easily solved to get an analytical solution in the contracting phase ($H<0$):
\begin{equation}
\label{contractsolG}
    \dot\phi\sim \frac{1}{\sqrt{2(t_0-t)}}~,~~~H=\frac{1}{2(t-t_0)}~.
\end{equation}
This indicates that this model will give a bounce solution with $w=1/3$ (radiation-like) in the contracting phase. In Fig.~\ref{Galileona} we plot the scale factor from the contracting phase to the expanding phase. 

\begin{figure}[htbp]
    \centering
        \includegraphics[width=0.6\textwidth]{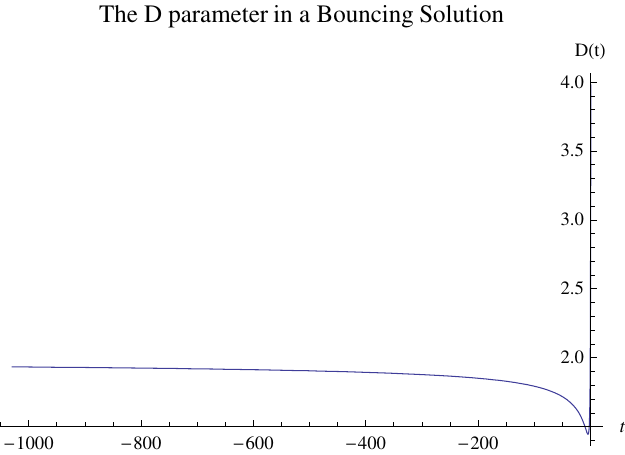}
    \caption{The evolution of the parameter $D$ with respect to cosmic time $t$ in a Galileon type quintom bounce. In the whole process, $D$ is shown to be keeping above $0$, indicating that there is no ghost instability. The figure is taken from Ref.~\cite{Qiu:2011cy}.}
    \label{GalileonD}
\end{figure}

To see how the ghost mode is eliminated in this kind of model, it is useful to apply the ADM formalism \cite{Arnowitt:1962hi} to write down the second order perturbed action. In this formalism, the perturbed metric is
\begin{equation}
    ds^2=-N^2dt^2+a^2e^{2\zeta}\delta_{ij}(dx^i+N^idt)(dx^j+N^jdt)~,
\end{equation}
where $N(x)$ and $N_i(x)$ are the lapse function and shift vector, while $\zeta$ is the curvature perturbation. Moreover, in uniform-$\phi$ gauge we set $\delta\phi=0$. Then the 2nd order perturbed equation for the curvature perturbation is \cite{DeFelice:2011zh}:
\begin{equation}
    S_2=3\int dtd^3x a^3DM_p^2\left[\dot\zeta^2-\frac{c_s^2}{a^2}(\partial_i\zeta)^2\right]~,
\end{equation}
where 
\begin{eqnarray}
    D&=&\frac{2M_p^4H^2+2(F/M)^6\dot\phi^6+(F/M)^3M_p^2\dot\phi^4}{2[M_p^2H-(F/M)^3\dot\phi^3]^2}~,\\
    c_s^2&=&\frac{-2M_p^4\dot H+2(F/M)^3M_p^2H\dot\phi^3-2(F/M)^6\dot\phi^6+6(F/M)^3M_p^2\dot\phi^2\ddot\phi}{3[2M_p^4H^2+(F/M)^3M_p^2\dot\phi^4+2(F/M)^6\dot\phi^6]}~.
\end{eqnarray}
From this result, we can see that the factor in front of the kinetic term of the perturbation $\zeta$ is positive definite, therefore there is no ghost mode. We also plot $D$ in the bounce solution in Fig.~\ref{GalileonD}.

However, as a bounce solution with a radiation-like contracting phase, it cannot give rise to the scale-invariant power spectrum required by the observational data. To remedy this defect, we introduce another curvaton field which has nonminimal coupling to the Galileon field to generate the power spectrum. A usual curvaton is a light field in an inflationary phase, which is used to give rise to a scale-invariant spectrum instead of the background field \cite{Lyth:2001nq, Moroi:2001ct, Cai:2011zx}.  But in a radiation-like contracting background, the curvaton needs to couple to the background field to make itself ``feel" that it was in an inflationary background. A possible action of the coupled curvaton is \cite{Creminelli:2010ba, Qiu:2011cy, Qiu:2013eoa, Qiu:2024sdd}
\begin{equation}
    S_\sigma=\int d^4x\sqrt{-g}[-\phi^q(\partial\sigma)^2]~,
\end{equation}
where $\sigma$ is the curvaton field, while $q$ is a parameter. The equation of motion for the curvaton is
\begin{equation}
\label{curveqG}
    \sigma''+\left(2{\cal H}+\frac{q\phi'}{\phi}\right)\sigma'-\nabla^2\sigma=0~.
\end{equation}
Since, in a radiation-dominant phase, from solution \eqref{contractsolG} one has $\phi\sim \eta$, ${\cal H}\sim \eta$, then the solution of Eq. \eqref{curveqG} (in momentum space) is
\begin{equation}
    \sigma_k\sim c_1 k^{-(1+q)/2}|\eta|^{-(1+q)}+c_2k^{(1+q)/2}
\end{equation}
in the super-horizon approximation. The power spectrum can be scale-invariant if:\\
1) $q=2$: 
\begin{equation}
    \sigma_k\sim c_1 k^{-3/2}|\eta|^{-3}+c_2k^{3/2}~,~~~P_\sigma\equiv\frac{k^3}{2\pi^2}|\sigma|^2\simeq \frac{c_1^2}{2\pi^2}|\eta|^{-6}~,
\end{equation}
or 2) $q=-4$:
\begin{equation}
    \sigma_k\sim c_1 k^{3/2}|\eta|^3+c_2k^{-3/2}~,~~~P_\sigma\equiv\frac{k^3}{2\pi^2}|\sigma|^2\simeq \frac{c_2^2}{2\pi^2}~.
\end{equation}
However, the equation of motion \eqref{curveqG} for the background gives the energy density for $\sigma$ field: $\rho_\sigma \sim (t_0-t)^{-3-q/2}$, while the radiation-like background scales as $\rho_{bg}\sim (t_0-t)^{-2}$. To obtain a stable evolution where the energy density of the curvaton field does not overwhelm the background and ruin the bounce, we require $q\geq -2$. Therefore the solution with $q=2$ is chosen.

Meanwhile, one can also calculate the tensor power spectrum in this model. Since the gravity part is not altered from Einstein's Gravity, the equation of motion for the tensor power spectrum is the same as the previous one, namely Eq. \eqref{perteqtensor}. For the radiation-dominated background $a\sim \eta$, one has
\begin{equation}
    P^T(k)\equiv \frac{k^3}{2\pi^2}\left|\frac{v_k}{a}\right|^2\sim k^2~,~~~n_T\equiv \frac{d\ln P^T(k)}{d\ln k}=2~.
\end{equation}
Therefore, a blue-tilted tensor power spectrum is obtained. However, it obviously depends on the contracting background as well as the gravity sector of the model. When the model is different, it is possible to obtain scale-invariant tensor spectrum as well.

\subsection{Bounce with modified gravity}
Bouncing cosmologies can also be realized in modified gravity models \cite{Nojiri:2017ncd}. In modified gravity, the Friedmann equations are no longer of the form~\eqref{friedmann}, while the modified gravity will act as effective $\rho$ and $p$.  Thus even in the absence of exotic matter, they can still effectively realize a quintom scenario and bounce the Universe. A large class of modified gravity models is called metric affine gravity, which contains other geometric quantities such as the torsion tensor \cite{Einstein:1929xrj, Unzicker:2005in, Hayashi:1979qx, Ferraro:2006jd, Bengochea:2008gz, Linder:2010py, Chen:2010va}:
\begin{equation}
    T^\alpha_{\mu\nu}=\Gamma^\alpha _{\nu\mu}-\Gamma^\alpha_{\mu\nu}~,
\end{equation}
whose existence means that the connection is no longer symmetric, and the nonmetricity tensor \cite{Nester:1998mp,
BeltranJimenez:2017tkd, BeltranJimenez:2019tme}:
\begin{equation}
    Q_{\alpha\mu\nu}\equiv\nabla_\alpha g_{\mu\nu}=\frac{\partial g_{\mu\nu}}{\partial x^\alpha}-g_{\nu\sigma}\Gamma^{\sigma}_{\mu\alpha}-g_{\sigma\mu}\Gamma^{\sigma}_{\nu\alpha}~,
\end{equation}
whose existence means that the covariant derivative of metric is no longer vanishing. Both of these tensors can be contracted into scalars:
\begin{eqnarray}
    T&\equiv&-\frac{1}{4}T^\rho_{\,\mu\nu}T^{\mu\nu}_{\,\,\rho}+\frac{1}{4}T^\rho_{\,\mu\nu}T^{\nu\mu}_{\,\,\rho}+\frac{1}{4}T^\rho_{\,\mu\nu}T^{\,\mu\nu}_{\rho}+\frac{1}{2}T^\mu_{\,\mu\nu}T^{\alpha\nu}_{\,\,\alpha}-\frac{1}{2}T^\nu_{\,\mu\nu}T^{\alpha\mu}_{\,\,\alpha}~,\\
    Q&\equiv&\frac{1}{2}Q_{\mu\nu\lambda}Q^{\lambda\mu\nu}-\frac{1}{4}Q_{\mu\nu\lambda}Q^{\mu\nu\lambda}+\frac{1}{4}Q^{\,\,\,\,\mu}_{\lambda\mu}Q^{\lambda\,\,\mu}_{\,\,\mu}-\frac{1}{2}Q^{\,\,\,\,\mu}_{\lambda\mu}Q^{\mu}_{\,\,\mu\lambda}~,
\end{eqnarray}
which are called the torsion scalar and the nonmetricity scalar, respectively. Moreover, people interestingly found that these two scalars have a simple relation to the Ricci scalar $R$:
\begin{equation}
    R=T-2\hat\nabla^\nu T^\mu_{\,\mu\nu}~,~~~R=Q-\hat\nabla_\alpha(Q^{\alpha\,\,\mu}_{\,\,\mu}-Q^{\mu\,\,\alpha}_{\,\,\mu})~,
\end{equation}
where $\hat\nabla$ denotes the covariant derivative with respect to the Christoffel symbol. For this reason, when these scalars act as the gravity sector of the Universe's action, in the minimal case they will be mathematically equivalent to each other \cite{Garecki:2010jj, Aldrovandi:2013wha, BeltranJimenez:2018vdo, BeltranJimenez:2019esp}:
\begin{equation}
    \int d^4x\sqrt{-g}R=\int d^4x\sqrt{-g}T=\int d^4x\sqrt{-g}Q~,
\end{equation}
and all of the physical laws originating from this action are the same. Thus the gravity theories of $T$ and $Q$ are equivalent to GR. However, when the actions are extended to arbitrary functions of $R$, $T$ and $Q$, they will be different from each other since the total derivative cannot be removed
\begin{equation}
     \int d^4x\sqrt{-g}f(R)\neq\int d^4x\sqrt{-g}f(T)\neq\int d^4x\sqrt{-g}f(Q)~.
\end{equation}
Then we will have different actions with different physical consequences \cite{Cai:2015emx, Heisenberg:2023lru, CANTATA:2021asi}. For example, when applied to FRW spacetime where the metric is $ds^2=-dt^2+a^2(t)\delta_{ij}dx^idx^j$, the Friedmann equations for the cases of $T$ and $Q$ should be:
\begin{eqnarray}
\label{MGeq1}
    6f_TH^2+\frac{1}{2}f&=&\rho_m~,\\
\label{MGeq2}
    (f_T-12H^2f_{TT})\dot H&=&-\frac{1}{2}(\rho_m+p_m)~,~~~(or~T\rightarrow Q )
\end{eqnarray}
where $T=Q=-6H^2$ \footnote{Note that in some literature, $Q$ is defined as $Q=6H^2$.  Then, in order to match with GR, the action of $f(Q)$ should be $S=-\int d^4x\sqrt{-g}f(Q)$ instead.}. Moreover, the Friedmann equations for $f(R)$ will be \cite{DeFelice:2010aj}:
\begin{eqnarray}
    3f_RH^2+3H\dot f_R-\frac{1}{2}f_RR+\frac{1}{2}f&=&\rho_m~,\\
    f_R\dot H+\frac{1}{2}\ddot f_R-\frac{1}{2}H\dot f_R&=&-\frac{1}{2}(\rho_m+p_m)~.
\end{eqnarray}

Various works on constructing bounce models in framework of modified gravity theories such as $f(R)$, $f(T)$ and $f(Q)$ has been done \cite{Carloni:2005ii, Cai:2011tc, deHaro:2012zt, Amoros:2013nxa, Bamba:2013fha, Haro:2014wha, Odintsov:2015uca, Bamba:2016gbu, Hohmann:2017jao, Qiu:2018nle, Bajardi:2020fxh, Mandal:2021wer, Hu:2023ndc}. For such theories, it is convenient to construct a bounce scenario by finding an Ansatz bounce solution first, then reconstructing the forms of the functionals. We consider a simple Ansatz
\begin{equation}
\label{MGa}
    a(t)=a_B(1+\alpha t^2)^{\frac{1}{3(\gamma+1)}}~,
\end{equation}
therefore the equation of state far from the bounce is $w=\gamma$. According to Eq. \eqref{MGa}, the Hubble parameter is given by
\begin{equation}
\label{MGhubble}
    H(t)=\frac{\dot a}{a}=\frac{2\alpha t}{3(1+\gamma)(1+\alpha t^2)}~.
\end{equation}
In principle, it is straightforward to substitute \eqref{MGhubble} into the Friedmann equations \eqref{MGeq1}, \eqref{MGeq2} to obtain the functional of $f$. However, the calculation process is difficult and we usually do not have an analytical solution over the whole period, but only have a numerical solution, or analytical solutions in certain periods, for which we refer the readers to e. g. Ref.~\cite{Cai:2011tc, Hu:2023ndc}. 

More detailed calculations and results can be found in Ref.~\cite{Cai:2011tc, Hu:2023ndc}, where realistic bounce models were obtained, and the evolutions of background and perturbations are discussed extensively. Although the action is different, at the leading order in perturbation theory, one can get similar equation of motion as in field theory models, and a nearly scale-invariant power spectrum can be obtained. See also \cite{Qiu:2010ch, Qiu:2010vk, Bamba:2014mya, Singh:2018xjv, Ilyas:2020qja, Agrawal:2021rur} for bounce models constructed from other modified gravity theories.

\subsection{Beyond a bounce: Cyclic scenarios}

A simple and precise example of a quintom cyclic model is given by the two-field action \cite{Xiong:2008ic}:
\begin{equation}
    S=\int d^4x\sqrt{-g}\left[\frac{1}{2}\partial_\mu\phi\partial^\mu\phi-\frac{1}{2}\partial_\mu\psi\partial^\mu\psi-V(\phi,\psi)\right]~,
\end{equation}
Thus the energy density and pressure of the universe are
\begin{equation}
\label{cyclicrhoandp}
    \rho=\frac{1}{2}\dot\phi^2-\frac{1}{2}\dot\psi^2+V(\phi,\psi)~,~~~p=\frac{1}{2}\dot\phi^2-\frac{1}{2}\dot\psi^2-V(\phi,\psi)~,
\end{equation}
while the equations of motion for the $\phi$ and $\psi$ fields are
\begin{equation}
\label{cycliceq}
    \ddot\phi+3H\dot\phi+V_{,\phi}=0~,~~~\ddot\psi+3H\dot\psi-V_{,\psi}=0~.
\end{equation}
To realize a cyclic scenario, we consider a coupled potential of the form
\begin{equation}
    V(\phi,\psi)=(\Lambda_0+\lambda\phi\psi)^2+\frac{1}{2}m^2\phi^2-\frac{1}{2}m^2\psi^2~.
\end{equation}
Due to the symmetry of the potential under the transformation: $\phi\rightarrow i\psi$, $\psi\rightarrow -i\phi$, the two scalar fields will dominate alternately.  Due to the couplings between two fields, the potential will be bounded from below. For Eq. \eqref{cycliceq} (with Hubble parameter satisfying the Friedmann equation and the energy density comes from Eq. \eqref{cyclicrhoandp}), we can obtain an Ansatz solution
\begin{equation}
    \phi=\sqrt{A_0}\cos mt~,~~~\psi=\sqrt{A_0}\sin mt~,
\end{equation}
where $A_0$ is the oscillation amplitude of the fields. Moreover, the Hubble parameter turns out to be:
\begin{equation}
\label{cyclichubble}
    H=\frac{\sqrt{3}}{3M_p}(\Lambda_0+\Lambda_1\sin 2mt)~,
\end{equation}
where we define $\Lambda_1=\sqrt{3}mA_0/(4M_p)$.

\begin{figure}[htbp]
    \centering
        \includegraphics[width=0.45\textwidth]{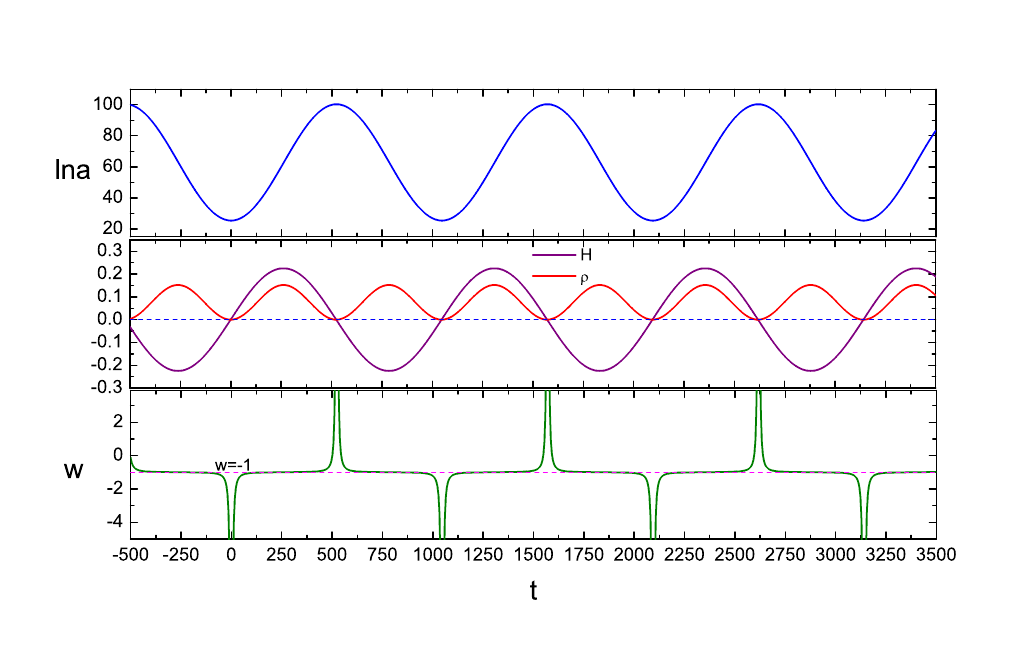}
        \includegraphics[width=0.45\textwidth]{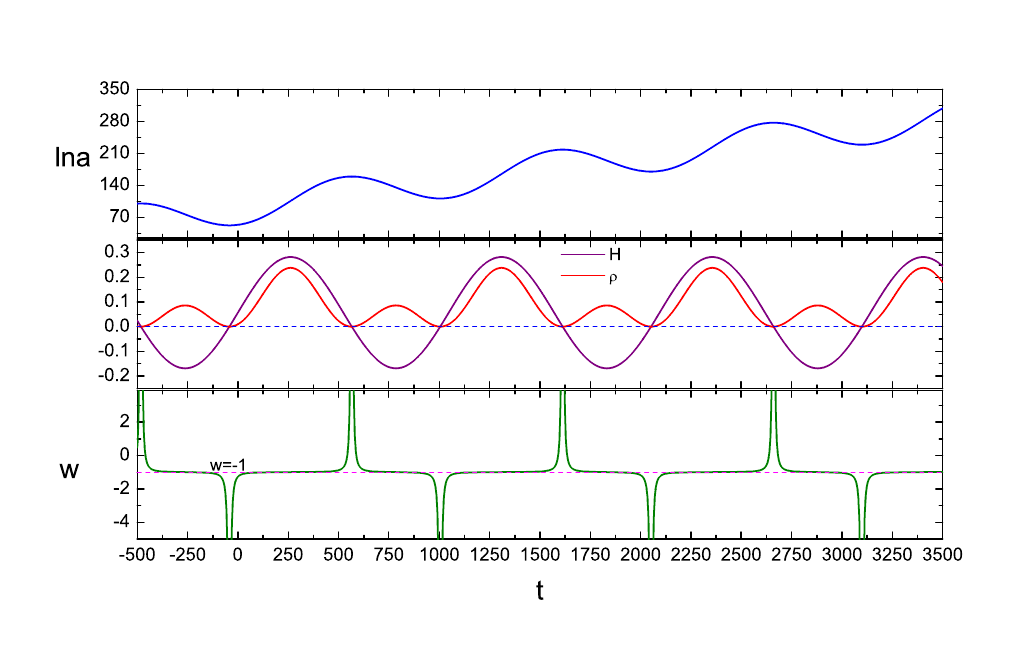}
        \includegraphics[width=0.45\textwidth]{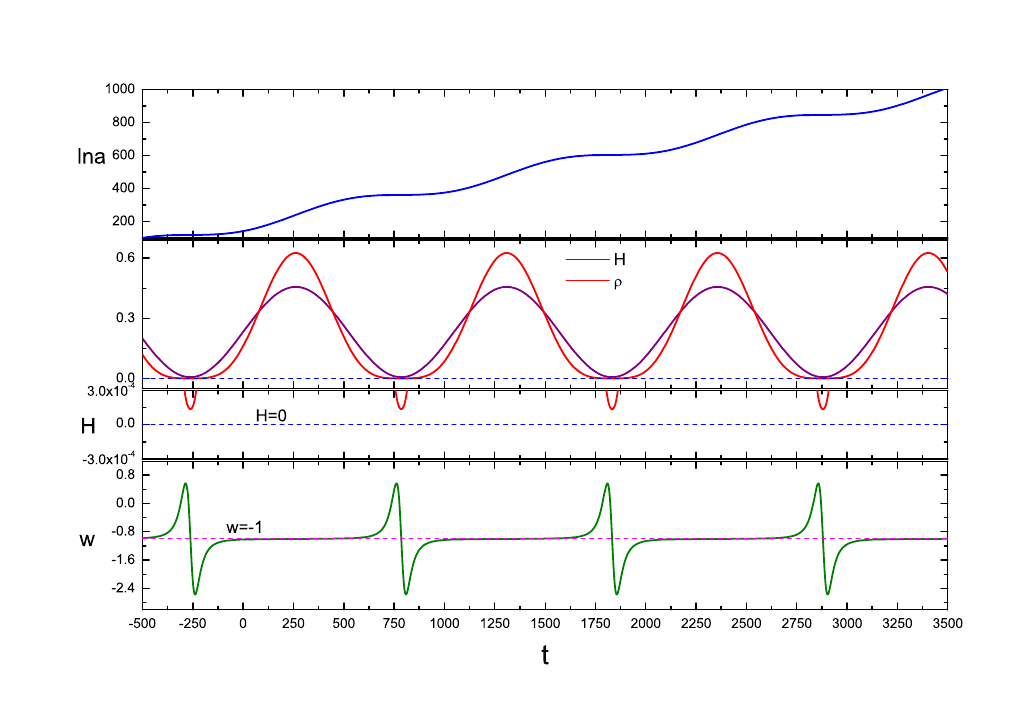}
        \includegraphics[width=0.45\textwidth]{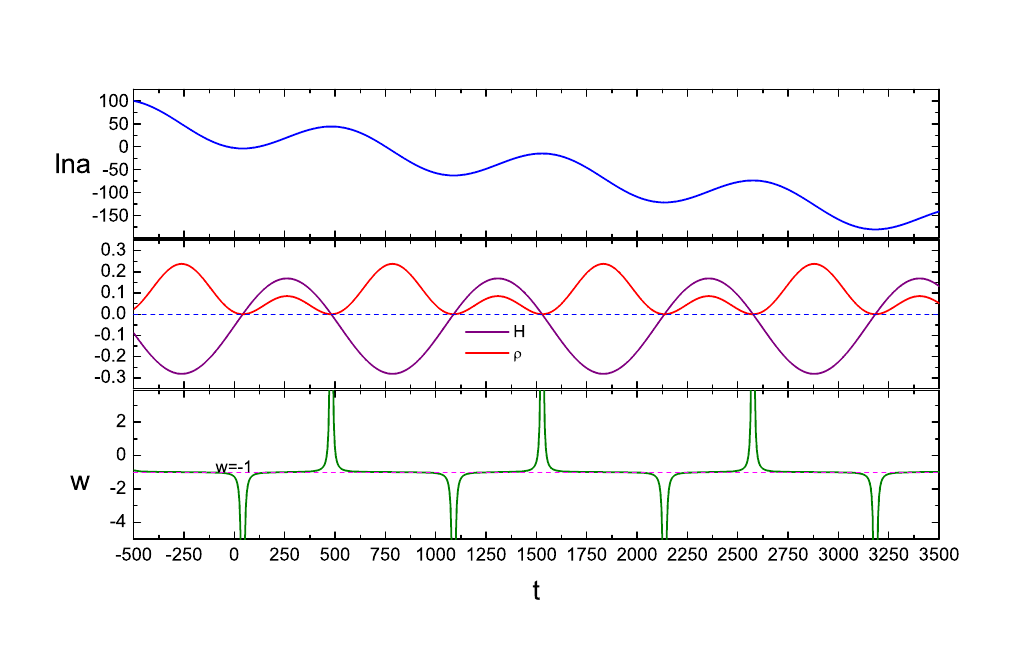}
    \caption{The evolution of $\ln a$, $H$, $\rho$ and $w$ with respect to cosmic time $t$, corresponding to blue, purple, red and green lines, respectively. In the numerical calculation, the parameters are taken to be $m=3\times 10^{-3}$ and $\Lambda_1\simeq 0.39$, and $\Lambda_0$ is taken as $0$ (left-top), $0.10$ (right-top), $0.402$ (left-bottom) and $-0.10$ (right-bottom) respectively, corresponding to the cases of $\Lambda_0=0$, $0<\Lambda_0<\Lambda_1$, $\Lambda_0\geq\Lambda_1$ and $-|\Lambda_1|<\Lambda_0<0$. The figures are taken from Ref.~\cite{Xiong:2008ic}.}
    \label{cyclicplot}
\end{figure}

From Eq. \eqref{cyclichubble} one can see that the Hubble parameter shows an oscillating behavior, and since $H=d\ln a/dt$, the scale factor can be oscillating as well. Moreover, different choices of the parameters $\Lambda_0$ and $\Lambda_1$ can give rise to different oscillating behaviors. In Fig. \ref{cyclicplot} we provide 4 possible kinds of oscillating behaviors of $\ln a$, $H$ and EoS $w$ given by Eq. \eqref{cyclichubble}. The first case is for $\Lambda_0=0$ where both $H$ and $\ln a$ are oscillating with an equal amplitude. This is a standard oscillating Universe, but it suffers from the entropy problem. The second case is for $0<\Lambda_0<\Lambda_1$. In this case, the center value of the Hubble parameter is positive, so the period of $H<0$ in each cycle will be shorter than that of $H>0$, therefore, the $\ln a$ is oscillating with the total amplitude increasing. The third case is for $\Lambda_0\geq \Lambda_1$. In this case the Hubble parameter is always larger than zero, so the Universe is always expanding with increasing $\ln a$. However, the expansion will be accelerating or decelerating alternately. Since there is no transition process from contraction to expansion, the EoS will also be regular all the time. The last case is for $-|\Lambda_1|<\Lambda_0<0$ where the center value of Hubble parameter is negative. In this case, period of $H<0$ in each cycle will be longer than that of $H>0$, therefore, the $\ln a$ will be decreasing in total, but have short periods of increase in each cycle. Therefore, it still has room to explain our current expanding universe.    

Apart from bouncing and cyclic scenarios, one should note that a non-singular Universe can also be realized via emergent behavior, when the Universe approaches a static model in the infinite past \cite{Brandenberger:1988aj, Ellis:2002we, Ellis:2003qz}. In this case, $H\rightarrow 0$ and $\rho\rightarrow 0$, while in order to enter into a realistic expanding phase, $H$ still needs to gradually increase, with $\dot H>0$. Thus the EoS of the Universe will go from $-\infty$ to above $-1$ as well. Therefore, in the emergent scenario, the Universe also realizes quintom-like behavior \cite{Cai:2012yf, Cai:2013rna}.


In conclusion, in this paper we briefly reviewed the Quintom model of dark energy and its applications to constructing bouncing cosmologies. Specifically, we have considered three classes of quintom scenarios: a two-field model, a single scalar with higher derivatives and modified gravity. 

Recently, other non-singular bounce scenarios making use of matter with EoS crossing $-1$ have been proposed. In Ref.~\cite{bbhp} Bardeen, Bars, Hanson and Peccei (BBHP) introduced the folded string, which is a classical solution of a limit of the Nambu model.  This was generalized by Itzhaki~\cite{1808.02259}, who found that in a classical (1+1)-dimensional string model with an increasing dilaton field, there are Instant Folded String (IFS) solutions.  These are essentially the BBHP folded strings, with the roles of time and space reversed so that they spontaneously nucleate\footnote{A quantum description of the IFS was proposed in Ref.~\cite{2209.04988}, allowing for a calculation of the IFS production rate.   An FZZT-brane~\cite{0001012} was used to cutoff the infrared divergence in the string energy, although in the Appendix the authors found that the brane suffers from several pathologies.}.
The IFS violates the null energy condition (NEC)~\cite{2101.10142}, a fact which was used to create a non-singular bouncing cosmology in Ref.~\cite{2412.02630}.  In Ref.~\cite{2508.09745}, Itzhaki, Peleg and Steinhardt showed that by introducing a particular potential for the dilaton, together with an extra field that becomes massless at an enhanced symmetry point \cite{0403001}, the bounce of Ref.~\cite{2412.02630} can be incorporated into a cyclic bouncing cosmology, complete with reheating performed by the extra field.


\section*{Acknowledgments}
TQ is supported by the National Key R$\&$D Program of China Grant No. 2021YFC2203100, and the National Natural Science
Foundation of China No.12575053. JE is supported by the Higher Education and Science Committee of the Republic of Armenia (Research Project No. 24RL-1C047). Yang Liu is supported by the National Natural Science
Foundation of China No.1240300.

\bibliographystyle{apsrev4-1}
\bibliography{References}

\end{document}